\documentclass{svmult}
\usepackage{natbib}
\usepackage{makeidx}
\usepackage{graphicx}
\usepackage{multicol}
\usepackage{url}

\newcommand{\enzo}{\code{Enzo}}
\newcommand{\code}[1]{\textsf{#1}}

\makeindex

\begin{document}
%
%--------------- Title Page ----------------
%
\title*{The Impact of AMR in Numerical Astrophysics and Cosmology}
\author{Michael L. Norman}
\institute{Laboratory for Computational Astrophysics at the 
Center for Astrophysics and Space Sciences, 
University of California at San Diego, La Jolla, CA 92093, USA}
%\texttt{bwoshea@cosmos.ucsd.edu}
\titlerunning{Impact of AMR}
\authorrunning{M. L. Norman}
\maketitle

%
%--------------------- Abstract ----------------------------
%
\textbf{Abstract.}  
I survey the use and impact of adaptive mesh refinement (AMR) simulations in numerical astrophysics and cosmology. Two basic techniques are in use to extend the dynamic range of Eulerian grid simulations in multi-dimensions: cell refinement, and patch refinement, otherwise known as block-structured adaptive mesh refinement (SAMR). In this survey, no attempt is made to assess the relative merits of these two approaches. Rather, the discussion focuses on how AMR is being used and how AMR is making a scientific impact in a diverse set of fields from space physics to the cosmology of the early universe. The increased adoption of AMR techniques in the past decade is driven in part by the public availability of AMR codes and frameworks. I provide a partial list of resources for those interested in learning more about AMR simulations. 

%
%--------------------- Introduction ------------------------
%
\section{Introduction}\label{sec:intro}

Since its introduction roughly 20 years ago \cite{Berger84}, adaptive mesh refinement (AMR) has emerged as an important class of numerical techniques for improving the accuracy and dynamic range of grid-based calculations for fluid dynamics problems. Such problems, especially compressible flow, develop steep gradients (shock waves and contact discontinuities) which, in the absence of mesh refinement, become sources of error for the global solution (e.g., \cite{Woodward84}). Through appropriate local mesh refinement, AMR can be thought of as a numerical technique for optimizing the quality of a numerical solution for a given computational cost (e.g., \cite{Berger89}). 

The last ten years have seen the application of AMR methods to problems in space physics 
%(de Zeeuw 1998, Groth et al. 1999, 2000a,b)
\cite{deZeeuw98,Groth99,Groth00}, astrophysics 
%(Falle & Raga 1993, 1995; Klein et al. 1994, 1998, 2003; 
%Duncan & Hughes 1994; Komissarov & Falle 1997, 1998; Truelove et al. %1997; Khokhlov 1998; Falle et al. 2002; Calder et al. 2002) 
\cite{Falle93, Falle95, Klein94, Klein98, Klein03, Truelove97, 
Truelove98, Khokhlov98, Kritsuk01, Falle02, Calder02},
and cosmology 
%(Bryan & Norman 1997, 2000; Kravtsov et al.1997; Kravtsov & 
%Klypin 1999; Abel, Byran & Norman 2000, 2002; Loken et al. 2002; 
%Tassis et al. 2003; Nagai & Kravtsov 2003). 
\cite{Bryan97, Bryan00, Kravtsov97, Klypin99, ABN00, ABN02, Loken02,
Tassis03, Nagai03}.
Here, a variety of physical processes may operate singly or together in astrophysical fluids to expand the range of important length- and time-scales. Such processes include gravity and gravitational instability, reaction kinetics, magnetic reconnection, radiation transfer, ionization fronts, etc. AMR has also been applied to the solution of gravitational N-body problems \cite{Couchman91, Kravtsov97}, and to hybrid particle-fluid simulations in cosmology \cite{Bryan97, Bryan99}. In such applications, the physics is intrinsically multi-scale, and AMR can be thought of as a numerical technique for extending the dynamic range of resolved physics, regardless of the computational cost. It is these sorts of applications which are reviewed here, and where AMR can make a scientific and not just an economic impact. 

AMR is also having a positive impact on the methodology of computational physics itself. I am referring to the validation of computational codes through resolution studies. With AMR, it is now practical on today's supercomputers to perform resolution studies over a sufficient range of scales to obtain convergent results on properties of interest which may be compared with laboratory experiments (e.g., \cite{Calder02}). 

Table 1 shows the impressive diversity of topics AMR has been applied to. Particularly interesting is the range of physical processes AMR has been adapted to model. This is listed in the second column of Table 1. As of today, AMR has been successfully applied to ideal gas dynamics (Newtonian and special relativistic), reactive gas dynamics, MHD (ideal and resistive), self-gravitating gas dynamics and MHD, N-body dynamics, and hybrid fluid/N-body systems. As we heard at this conference, methods are under development for radiative transfer (Howell), radiation hydrodynamics (Weaver), and solid mechanics (Falle). It is clear AMR is a method of wide applicability, and one that is growing in its impact. This growth is fueled in part by the public availability of AMR codes and frameworks (see Section 3).

\begin{table}
\centering
\begin{tabular}{|l|l|l|}
\hline
\bf{Topic}   & \bf{Physics} & {\bf Select References} \\
\hline
\hline
Code validation & HD, reactive HD & \cite{Calder02} \\
\hline
Solar and space physics & MHD & \cite{Groth99} \\
\hline
Supernovae and nucleosynthesis & reactive HD & \cite{Kifonidis00, Timmes00, Gamezo03} \\
\hline
Interstellar medium & HD, MHD & \cite{Klein94, Kritsuk01, Cid96, Balsara01} \\
\hline
Star formation & grav HD, grav RHD & \cite{Truelove98, Klein03} \\
\hline
Astrophysical jets & HD, rel HD & \cite{Falle95, Hughes97, KF97} \\
\hline
N-body dynamics & particles, grav & \cite{Kravtsov97, Klypin99} \\
\hline
Hydrodynamic cosmology & hybrid & \cite{Bryan97, Nagai03} \\
\hline
\hline

\end{tabular}
\newline
\caption{Classes of AMR applications.}
\label{applications}
\end{table}
\medskip

In this paper I survey the use and impact of adaptive mesh refinement simulations in numerical astrophysics and cosmology. Two basic techniques are in use to extend the dynamic range of Eulerian grid simulations in multi-dimensions: cell refinement (CR), and patch refinement (PR), otherwise known as block-structured adaptive mesh refinement (SAMR). Details of these two approaches are given elsewhere in this volume. In this review, no attempt is made to assess the relative merits of these two approaches. Rather, the discussion focuses on how AMR is being used and how AMR is making a scientific impact in a diverse set of fields from space physics to the cosmology of the early universe. At the end, I provide a partial list of software resources for those interested in learning more about AMR simulations. [{\bf NB.} The following is meant to be representative rather than complete. I apologize to authors in advance if their work is not mentioned.]

%
%--------------------- Applications ------------------------
%
\section{Applications of AMR}\label{sec:apps}%
\subsection{Solar and Space Physics}
Coronal mass ejections (CMEs) are transient solar events in which mass and magnetic field are ejected from the solar surface. These dynamic events originate in closed magnetic field regions of the corona. They produce large-scale reconfigurations of the coronal magnetic field and generate solar wind disturbances which manifest as geomagnetic storms here on Earth. Groth et al. \cite{Groth99, Groth00} have applied AMR to 3D simulations of solar coronal outflows and ejections. The central question which motivates their research is what is the mechanism and timescale for coronal mass ejections (CMEs)? The simulations solve the equations of ideal MHD in 3D Cartesian geometry, supplemented with the Sun's gravitational field and a coronal heating term. The equations are solved using the upwind, cell-centered, finite volume scheme of Powell \cite{Powell94, Powell95}, which is a Godunov-type scheme for ideal MHD. Numerical fluxes are computed using the approximate Riemann solvers of both Roe \cite{Roe81} and Harten et al. \cite{Harten83}, adapted to the MHD eigensystem. The MHD solver is married to the parallel AMR framework of de Zeeuw et al. \cite{deZeeuw98}. In this approach, the base grid is decomposed into blocks of constant size, each of which is assigned to a separate processor. If any cell within a block are flagged for refinement, the entire block is refined by a factor of two in each dimension, resulting in eight sub-blocks of the size of the parent block. 

The CME problem is initialized with a model of the quiescent solar wind which involves a high speed wind and open magnetic field lines at high solar latitudes, and a lower wind speed and closed magnetic field lines in the equatorial region. The CME event is triggered with the introduction into the computational domain of a density pulse in the closed field region. The mass loading inflates the closed field region until it bursts open. Uisng AMR, they are able to follow the CME outburst to $\frac{1}{2}$ AU.  According to the authors, the benefits of AMR in this demonstration calculation are cost/memory savings, and the ability to refine a region of interest (the magnetic X-point) which moves through the computational volume. Although they were unable to answer the ultimate question, the authors argue that AMR combined with more realistic solar magnetic field configurations and initiation mechanisms will lead to improved understanding of CMEs. 
\subsection{Supernovae and Nucleosynthesis}
An important and growing class of applications for AMR simulations is calculating the mechanisms and chemical yields of supernovae models in three dimensions. In recent years, the importance of convective/turbulent motions in both Type Ia (thermonuclear) and Type II (core collapse) supernovae has been recognized \cite{Khokhlov95, Herant94}. Fully 3D simulations are thus required, and AMR serves two fundamental roles. The first is reducing memory and cpu requirements for expensive 3D simulations, and the other is resolving and tracking dynamic interfaces where energy generation and nucleosynthesis takes place. 

Type Ia supernovae are believed to result from the explosive burning of carbon and oxygen in a white dwarf (WD) accreting matter from a binary companion. As the WD's mass approaches the Chandrasekhar limit, a small mass increase causes a substantial contraction of the star. The compression raises the temperature and thereby thermonuclear reaction rates, liberating energy which raises the temperature yet further. Because the WD is degenerate, a thermonuclear runaway ensues until thermal pressure becomes comparable to the degenerate electron pressure. At this point, the WD begins to expand, but not so fast that the thermonuclear reactions are quenched. According to models \cite{Reinecke02, Gamezo03}, a nuclear burning front propagates outward from the center of the expanding WD liberating of order $10^{51}$ ergs of energy---sufficient to unbind the star. 

Gamezo et al. \cite{Gamezo03} have developed a 3D AMR code to simulate the physics described above, and have addressed several key questions: (1) How does a degenerate C/O white dwarf explode and with what energy? (2) What is the nature and structure of the burning front? And (3) what fraction of the WD is burned? They solve Euler's equations of gas dynamics for four reactive species coupled to a nuclear reaction network. The energy equation includes nuclear energy generation and neutrino losses, as well as electron thermal conduction. A key additional incredient to the physical model is a flame capturing technique which adds one additional PDE to be solved for a reaction progress variable. The system is evolved for one octant of the WD star on a 3D Cartesian grid using a cell-refinement AMR technique called FTT (Fully Threaded Tree) developed by Khokhlov \cite{Khokhlov98}. The benefit of cell-refinement over SAMR in this application is that the region of interest is a surface---the flame front---which is more efficiently captured with a cell-refinement technique \cite{Khokhlov98}. 

Gamezo et al. find that the nuclear flame front is a deflagration front, meaning it advances subsonically into the unburned gas. The flame front is Rayleigh-Taylor (hereafter, RT) unstable and becomes highly convoluted, resembling the head of a cauliflower. AMR is used to track this highly distorted surface as it advances into the unburned star. They find that a healthy explosion results, but that a substantial fraction of the WD remains unburned at disassembly. This is contrary to observations, which are found to be more consistent with detonating WD models (e.g., \cite{Arnett94, Hoeflich95}). Gamezo et al. speculate that the turbulent flame may trigger a detonation at some point, which would complete the burning and remove this discrepancy. The physics of deflagration to detonation transition (DDT) depends on resolving the formation of ``hot spots" behind the turbulent flame front on scales comparable to the flame front thickness \cite{Khokhlov99}. This is tiny compared to the radius of the WD, and thus not captured by global simulations despite the use of AMR. At present DDT must be studied using simulations in small volumes (``subscale simulations".) 

An example of this is the work of Timmes et al. \cite{Timmes00}, who used the FLASH code \cite{Fryxell00} to study the cellular structure of carbon detonation waves in degenerate WDs. Their goal was to assess the effect of numerical resolution on the size and shape of the detonation cells and the shock wave interactions that create them. FLASH is a block-structured AMR code combining the piecewise parabolic method for gas dynamics, nuclear reactions, and the PARAMESH AMR library \cite{Fryxell00}. The simulation was done in a 2D column of gas with conditions representative of the WD. The detonation was initiated by sending a shock wave into the column with strength consistent with a steady 1D detonation wave (``CJ wave"). AMR was used to refine by up to a factor of 8 in cell size the reaction zone behind the shock. AMR thus provided the ability to have high resolution just in the detonation wave and track it as it propagated many times its width. 

Timmes et al. were able to converge on the size and shape of the detonation cells with rather modest resolution (about 20 cells per burning length scale), but found that the size and distribution of pockets of unburned fuel was very sensitive to resolution. As long as this scale is small compared to the WD radius, observations would average over these compositional differences. However, the burning length scale and hence detonation cell size becomes comparable to the WD radius as density drops toward the edge of the star \cite{Khokhlov91}. If the spectra of Type Ia supernovae reflect large scale abundance variations due to incomplete combustion, the results of Timmes et al would help define the minimum resolution requires to converge on the inhomogeneities. 

In a related work, Zingale et al. \cite{Zingale01} have applied FLASH to the dynamics of helium detonation on the surface of a neutron star in 2D. In this application, as in Gamezo et al., AMR provides the ability to resolve and track the nuclear burning front in an expanding envelope of gas.   

AMR is also making an impact in the understanding of the iron core collapse supernova explosion mechanism and subsequent explosive nucleosynthesis. Supernovae classified by observers as Type Ib and II are believed to be powered by the copious flux of neutrinos emitted as the iron core of a massive star implodes to form a neutron star or black hole \cite{Woosley02}. Earlier 2D simulations showed that neutrino heating sets up convective motions in material accreting onto the proto-neutron star/black hole, breaking spherical symmetry 
\cite{Herant94, Burrows95}. The consequence of this is that the shell of 56Ni which forms just outside this convective region and is later ejected by the supernovae is highly perturbed \cite{Kifonidis00}. 

AMR has not yet been applied to the very difficult problem of neutrino transport/heating in a 3D convective flow, although a large collaboration is attempting this \cite{Mezzacappa02}. Rather, Kifonidis et al. \cite{Kifonidis00, Kifonidis03}  have used AMR to simulate the evolution of the lumpy nickel shell as it is ejected by the explosion. AMR is used to track the expansion of the shell over a factor of 100 in radius and its fragmentation by the RT instability while maintaining high resolution in the shell. The authors map the results of a 2D axisymmetric core collapse simulation 30 ms after bounce into AMRA, a 2D AMR hydro code developed by Plewa \& M{\" u}ller \cite{Plewa99}. The equations of multispecies reactive hydrodynamics are solved on an adaptive spherical polar grid using the PPM algorithm \cite{Colella84} and the Consistent Multifluid Advection (CMA) scheme of Plewa \& M{\" u}ller. The latter minimizes numerical diffusion of nuclear species while preserving local mass conservation. Kifonidis et al. find that the 56Ni and other newly formed iron group elements are distributed throughout the inner half of the helium core by RT instabilities operating at the (Ni+Si)/O and (C+O)/He shell interfaces seeded by perturbations from convective overturn during the early stages of the explosion. Using AMR, they are able to carry the calculation out to 20,000 sec post-bounce, follow the details of RT growth and mixing. The result is that fast-moving clumps of 56Ni are formed with velocities up to ~4000 km/s. This offers a natural explanation for the mixing required in light-curve and spectroscopic modeling of Type Ib explosions, including SN1987a.
\subsection{Interstellar Medium}
The essential complexity of the interstellar medium (ISM) stems from the fact that it is not in equilibrium, dynamically or thermodynamically. A variety of heating and cooling processes operate to split the ISM into multiple thermal phases, each with their own characteristic densities, temperatures, and evolutionary timescales. Self-gravity concentrates the cold, dense phase into molecular clouds, which birth new stars. The most massive stars create large amplitude disturbances in the local heating rate through supernova shocks and ionization fronts, changing the dynamical and thermodynamical state of the gas. To simulate the ISM is rather akin to simulating the Earth's weather, which is affected by both local and global influences, and exhibits complexity in space and time on a vast range of scales. With AMR simulations, we may eventually be able to build an integrated model of the ISM which captures both its structural and statistical properties. At present, researchers are looking at individual processes in greater detail than ever before that ultimately may become part of an integrated model.

The propagation of interstellar shock waves in the ISM has received considerable attention because of their central role as a source of heat and momentum to the interstellar gas. The phenomenology is rich because the ISM is inhomogeneous, and the shocks themselves are typically radiative. A set-piece calculation is the interaction of a strong planar shock wave striking an isolated interstellar cloud, idealized as homogeneous and spherical. The interaction results in the compression and ultimate shredding of the cloud by a combination of Richtmyer-Meshkov, Rayleigh-Taylor, and Kelvin-Helmholtz instabilities. Since these instabilities all grow on the shock-accelerated cloud-intercloud interface, it is important that the interface be tracked with high resolution. 

The first AMR simulations were performed by Klein, Mc Kee \& Colella 
\cite{Klein94}. They wanted to calculate how long a shocked cloud would remain intact before mixing into the ISM. They solved the gas dynamical problem in 2D using block-structured AMR and a second-order Godunov hydrodynamics scheme as described in \cite{Berger89}. They found that for strong incident shocks, the evolution is determined by a single parameter: the ratio of cloud to intercloud densities. Using AMR, they performed a resolution study to determine the minimum resolution needed to capture the most destructive modes of the instabilities. They found that a minimum of 100 cells per cloud radius are needed with their second order-accurate scheme, and that the cloud is totally fragmented.

A closely related problem is the long-term evolution of the dense, metal-rich clumps ejected by supernova explosions (cf. \cite{Kifonidis03}). Because dense clumps are decelerated less than the diffuse interclump ejecta, they catch up to the supernova remnant (SNR) shell and puncture it from behind. The clump first encounters the reverse shock, traverses the high pressure intershock region, and then exits the forward shock if it survives to encounter the onrushing ISM. Cid-Fernandes et al. \cite{Cid96} used the 2D AMR code AMRA \cite{Plewa99} to simulate the evolution of a single clump with properties appropriate to a Ni clump from a core collapse SN. They initialized the freely expanding envelope of a Type II SN in spherical polar coordinates, assuming axisymmetry. They placed a clylindrical plug of denser material just upstream of the reverse shock, and followed its subsequent evolution by solving the equations of ideal gas dynamics including equilibrium radiative cooling. Three levels of mesh refinement tracked the blob, providing local resolution equivalent to a uniform grid of 1536 x 160 grid cells. The benefit of AMR over a uniform grid was a cost savings of 350\%. They found that the clump is strongly compressed in the intershock region by virtue of the high pressure and strong radiative cooling in the clump. While lacking the resolution of Klein et al. \cite{Klein94}, they concluded the clump would most likely be disrupted into secondary fragments before reaching the dense outer shell. They suggested that x-ray flares would result if the largest of these secondary clumps survived to strike the dense shell.

Two other AMR-enabled simulations of interstellar shock waves illustrate the richness of the phenomena. In the first, Poludnenko, Frank \& Blackman \cite{Poludnenko02} simulated the propagation of a planar, adiabatic shock through a clumpy medium. The motivation for the simulations was to understand mass-loading and mixing of stellar and galactic outflows by inhomogeneities in the ambient medium. A parameter survey was carried out to assess the effects of clump mass and spatial distribution on the flow. 2D AMR simulations where performed in planar geometry using the AMRCLAW package of Berger \& LeVeque \cite{Berger98}, which combines SAMR with a second order-accurate Godunov scheme for ideal gas dynamics. The contribution of AMR was to achieve high resolution in each of many clumps scattered randomly throughout the volume as they are shocked and sheared. Resolution equivalent to a 800 x 1600 uniform grid was achieved in the clumps for a fraction of the cpu and memory cost, which is particularly important when conducting a large parameter survey. They found that a critical longitudinal and transverse separation between clumps exists such that for $d < d_{crit}$ and $L< L_{crit}$, the post-shock flow is strongly interacting, leading to enhanced turbulence and mixing. 

In the second, a new type of instability was discovered in radiative shocks by Walder and Folini \cite{Walder98}. Radiative shock waves are found in many types of classical nebulae, like supernova remnants, planetary nebulae, Wolf-Rayet ring nebulae, etc. Typically, a shell of dense material is formed by the interaction of a fast outflow with the circumstellar medium (CSM) or interstellar medium (ISM). This shell will be bounded by two shocks: an outer, forward shock compressing the CSM/ISM, and an inner, reverse shock compressing the ejecta. Typically, the outer shock is radiative, while the inner shock is not because of the relative densities involved. The shocked media are separated by a contact discontinuity which is normally RT stable at late times because the shell decelerates. 

However, it is well known that strongly radiative shocks suffer from an overstability such that the shock oscillates about its steady-state position in the rest frame of the contact discontinuity \cite{CI82}. In multidimensions, Walder and Folini showed that different sections of the radiative shock oscillate independently, creating lateral pressure disturbances within the high pressure shell. These chaotic perturbations sometimes accelerate the CD and thereby excite the RT instability. This leads to fingers and clumps of dense, shocked CSM/ISM efficiently mixing with the shocked ejecta. It is suggested by the authors that the mixing will boost the X-ray emission, contribute to rapid variability in the emission spectra, and may contribute to the the clumpy/filamentary appearance of the nebulae mentioned. The numerical simulations were carried out in 2D using AMRCART, which combines block SAMR with a second order accurate Godunov solver for gas dynamics. Optically thin radiative cooling was included assuming equilibrium ionization. The simulation was carried out in the rest frame of the CD, and hence the role of AMR was not shock tracking, but rather maintaining high resolution near the unstable interface.
\subsection{Star Formation}
The gravitational collapse of a gas cloud to form a star is a notoriously difficult problem because of the large range of length- and time-scales that need to be resolved. This problem has been solved in one dimension assuming spherical symmetry using moving adaptive meshes and implicit time integration \cite{Winkler80a, Winkler80b}. Important structures that need to be resolved span some 9 decades in radius from inside the hydrostatic protostellar core to the edge of the accreting cloud. Important timescales vary by 6 decades from the sound crossing time in the hydrostatic core to the accretion time. AMR holds forth the promise that simulations covering this range of scales will become possible in 3D, permitting a self-consistent investigation of disk accretion and the dynamical role of magnetic fields. While this is still out of reach, important strides have been made on the early stages of non-spherical cloud collapse by Richard Klein and colleagues \cite{Truelove97, Truelove98, Klein98, Klein03} . They are investigating the important problem of binary star formation using adaptive mesh techniques. They calculate the gravitational fragmentation of slowly rotating molecular cloud cores in 3D, assuming the gas is isothermal and non-magnetic. The equations of isothermal, self-gravitating gas dynamics is solved using a second order-accurate Godunov scheme on a block SAMR grid. The Poisson equation for the gravitational potential is solved using multigrid relaxation as described in \cite{Truelove98}.

An important early finding was that the cloud would fragment artificially due to numerical perturbations unless the local Jeans length is resolved by at least 4 cells at all times \cite{Truelove97}. Since the Jeans length scales as $\rho^{-1/2}$, if collapse raises the central density by a factor $x$, then the grid spacing must be decreased by a factor $x^{-1/2}$ in order to avoid artificial fragmentation. The simulations of Truelove et al. \cite{Truelove97, Truelove98} used AMR was to follow compressions of $x \sim 10^8$ at which time the isothermal assumption breaks down. Local mesh refinement of up to a factor of $10^4$ in resolution was accomplished by recursively adding grids refined by a factor of 4 wherever the Jeans condition 
$J=\Delta x/\lambda_J <1/4$ was about to be violated. Simulations with fixed dynamic range X will inevitably violate the Jeans condition when $x>X$. Truelove et al.
 showed that an unperturbed cloud will fragment artificially shortly after the Jeans condition is violated \cite{Truelove98}. They went on to show that a slowly rotating cloud with an m=2 nonaxisymmetric perturbation does not fragment into a binary system as had previously been reported on the basis of fixed resolution simulations by Burkert \& Bodenheimer \cite{Burkert96}, but rather forms a singular isothermal filament in accord with analytic predictions \cite{Inutsuka92}. 

At sufficiently high number densities, however, the isothermal assumption breaks down because the cloud becomes optically thick to its own cooling radiation \cite{Larson03}. Boss et al. \cite{Boss00} simulated the non-isothermal evolution of the singular filament using both fixed as well as AMR grid codes. Opacity effects were modeled by adopting a barotropic equation of state with a variable gamma-law. The codes agreed providing the Jeans condition was obeyed in the fixed grid run. They found that the filament fragments into a binary system whose properties depend sensitively on the equation of state. Since the EOS only mocks up radiation trapping in an approximate way, the implication is that radiative transfer needs to be modeled self-consistently in future AMR simulations of cloud fragmentation. A flux-limited radiation diffusion algorithm for AMR grids has recently been introduced by Howell \& Greenough \cite{Howell03} which is beginning to be applied to star formation \cite{Klein03}.
\subsection{Astrophysical Jets}
Astrophysical jets are highly collimated, high speed bipolar outflows powered by disk accretion onto compact, gravitating objects. They manifest in a surprising diversity of systems and length scales, ranging from the pc-long optical jets from young stars \cite{Reipurth01} to the Mpc-long radio jets from active galactic nuclei and quasars \cite{Ferrari98}. Regardless of their origin, the jets themselves are interesting dynamically and morphologically because they sample both the deep potential wells where they were launched, and the interstellar/intergalactic medium they propagate through. Because of their high degree of collimation, the jets are believed to be hypersonic, and in the case of extragalactic jets, relativistic. A near universal feature of astrophysical jets, whether from stars or galaxies, is the occurrence of emission knots. Emission knots are patches of high emissivity arrayed along the length of the jet. These are generally interpreted as internal shock waves in the jet excited by internal or external perturbations \cite{Rees78}. 

Beginning in the early 1980s, there have been extensive numerical studies of the structure and dynamics of astrophysical jets with the framework of ideal gas dynamics and MHD. From the standpoint of simulations, the salient difference between protostellar jets and extragalactic jets is that the former are dense enough to be strongly radiative, whereas the latter are effectively adiabatic. One of the first applications of AMR in astrophysics was by Falle \& Raga \cite{Falle93, Falle95}, who studied the detailed structure of an emission knot in a radiative protostellar jet. 1D models by Raga et al. \cite{Raga90} showed that variations in the outflow velocity would create forward-reverse double shock pairs which propagate down the jet axis with the mean flow velocity, in good accord with observations. However, these models could not capture the bow shock appearance of the knots caused by the lateral expansion of the shocked gas into the ambient medium. Falle \& Raga \cite{Falle95} simulated the multidimensional structure of a single knot in the rest frame of the knot. The contribution of AMR was to resolve the strong cooling region and ionization structure behind the radiative shocks, which can be a small fraction of the jet radius. It is impractical to do this with uniform grids. The calculations were performed in 2D assuming axisymmetry using a block structured AMR grid. Six levels of grid refinement were used for an effective grid resolution of 1280 $\times$ 640 cells. The physics included gas dynamics, solved using the second-order Godunov scheme of Falle \cite{Falle94}, non-equilibrium ionization, and radiative cooling. The simulations showed that the knots could survive as coherent entities for many jet radii, and were morphologically similar to those observed. The ability to resolve the ionization structure with AMR allowed them to make synthetic emission maps in the commonly detected [S II] doublet and thereby diagnose the physical conditions in observed jets. 

Radio observations of radio-loud active galactic nuclei (AGN) mapped with VLBI techniques reveal one-sided jets with a stationary core and knots of emission that sometimes move superluminally \cite{Ferrari98}. This is conventially interpreted as the result of relativistic Doppler boosting and time dilation when observing a relativistic jet at small inclination angles \cite{BK79}.  With the advent of good algorithms for relativistic hydrodynamics \cite{Duncan94, Marti95, KF97},
%(e.g., Duncan & Hughes 1994, Marti, Muller & Ibanez 1994, 
%Komissarov & Falle 1996)
it becomes possible to model these sources. The structure, stability, and radio morphology of relativistic jets in compact extragalactic radio sources has been studied using AMR simulations by Hughes and collaborators \cite{Hughes97, Hardee98, Rosen99}.
%(Mioduszuski, Hughes & Duncan 1997; Hardee et al. 1998, 
%Rosen et al. 1999). 
Given the complexities of underlying hydro and the relativistic radiative transfer, they simply wanted to know whether such models resemble the data. They carried out 2D axisymmetric simulations using the code of Duncan and Hughes \cite{Duncan94}, which combines a second-order Godunov solver for the relativistic gas dynamics with the block-structured AMR method of Quirk \cite{Quirk91}. The contribution of AMR to this work is to crisply capture the internal shocks without resort to large uniform grids and supercomputers. Assuming a simple relation between synchrotron emissivity and gas pressure, Mioduszuski et al. computed synthetic radio maps for a variety of inclination angles, Lorentz factors, and Mach numbers. They found that the VLBI maps of superluminal sources are reasonably well fit with strongly perturbed "pulsed" relativistic jets seen nearly end-on. They found that temporal changes of the models' radio appearance is not easily related to the underlying hydrodynamic quantities due to differential Doppler boosting.
 
\subsection{Galaxies and Cosmology}
Although AMR was invented for accurately integrating the hyperbolic partial differential equations of fluid dynamics, the adaptive mesh used in conjunction with the particle-mesh (PM) and P$^3$M N-body techniques \cite{HE88} is extremely powerful for simulations of collisionless, self-gravitating systems of particles as arise in galactic dynamics and cosmological structure formation. AMR N-body codes have been developed by Couchman \cite{Couchman91}, Jessop et al. \cite{Jessop94}, Kravtsov et al. \cite{Kravtsov97}, Bryan and Norman \cite{Bryan97, Bryan00} and Knebe et al. \cite{Knebe01}. These codes differ in AMR data structures and how AMR is used to optimize the N-body calculation. 

Couchman  adaptively introduces one or two levels of block mesh refinement around highly clustered regions of particles in order to reduce the number of particle-particle pairs in a P$^3$M calculation of dark matter clustering, resorting instead to the faster PM calculation on the subgrids. The Poisson equation is solved on each level of the grid hierarchy using Fourier techniques with special Greens functions. The algorithm is referred to as AP$^3$M (Adaptive P$^3$M). 

Jessop et al. combine the classic PM scheme with block structured local mesh refinement to achieve high force resolution in condensed systems. The Poisson equation is solved at all levels of the grid hierarchy using Neumann boundary conditions interpolated from parent grids and an ADI relaxation scheme. The algorithm is referred to as PM$^2$ (Particle Multiple Mesh). 

Kravtsov et al. combine the octree cell refinement approach of Khokhlov \cite{Khokhlov98} with PM to create the ART (Adaptive Refinement Tree) code. The Poisson equation is solved using successive multilevel relaxation. 

Bryan \& Norman adapted and generalized Couchman's AP$^3$M to an arbitrarily deep AMR grid hierarchy and married it to a PPM-derived hydro solver (see next section.)  Two Poisson solvers are implemented: one based on Fourier techniques, and another using multigrid relaxation techniques (see O'Shea et al., these proceedings).

Knebe et al.'s code is similar to the ART code in that cell refinement and multigrid relaxation is used, but the underlying data structures and timestepping schemes are somewhat different.

Klypin et al. \cite{Klypin99} have applied the ART code to the gravitational clustering of cold dark matter in cosmological simulations of structure formation. The key difficulty of all such calculations is to resolve the scales on which galaxies form (1-10 kpc) in cosmological volumes large enough to sample the longest perturbation waves or to get good statistics ($\sim$ 100 Mpc). The required spatial dynamic range is therefore $10^4-10^5$ in 3D for multiple centers of interest (galaxies). AMR is one option for achieving such resolution; meshless tree codes are another (cf. \cite{Moore96}). Klypin et al. investigated the long-standing overmerging problem, in which N-body simulations of the formation of clusters of galaxies yield too few galaxy-sized dark matter (DM) halos compared with observations \cite{White76, Moore96}. Rather, early simulations found that the galaxy DM halos merged with one another as they orbited within the cluster potential. Originally, it was thought that overmerging was a consequence of the omission of dissipative baryons from the models (e.g., \cite{KHW92}). 

Klypin et al. showed that the overmerging problem is primarily a numerical resolution problem. Namely, that with inadequate force resolution, galaxy DM halos are numerically smeared out. As a consequence, the portion of the DM halo that is beyond its tidal radius gets stripped by cluster tidal forces as well as through close encouters with other galaxy halos. The DM halos essentially evaporate in a few orbits and their cores sink to the center of the cluster by dyamical friction. Klypin et al. combined analytic estimates and ART simulations to determine the resolution requirements to avoid overmerging. The simulations used $128^3$ DM particles and a base grid of $256^3$ cells in a volume 15 h$^{-1}$ Mpc on a side. Up to 7 levels of 2x cell refinement were permitted, for a maximum dynamic range of 32,000 and spatial resolution of 0.5 h$^{-1}$ kpc. They found that a force resolution of 1-2 h$^{-1}$ kpc and a mass resolution of $\sim 2 \times 10^8 h^{-1} M_{\odot}$ is sufficient to sample the population of galaxy DM halos in a rich cluster of galaxies.
\subsection{Hydrodynamic Cosmology}
The marriage of a gravitational N-body code for the cold dark matter in the universe with a hydrodynamics code to model the baryonic component is referred to as a cosmological hydrodynamics code. A natural marriage is a PM N-body code with an Eulerian gas dynamics code, and a number of such codes have been developed 
%(Cen et al. 1990, Yuan, Centrella & Norman 1991, Cen 1992, 
%Ryu et al. 1993, Anninos, Norman & Clarke 1994, Bryan et al. 1995, 
%Ricker, Dodelson & Lamb 2000). 
\cite{Cen92, Ryu93, Anninos94, ANC94, Bryan95, Ricker00}.
The spatial resolution of these codes is limited to the grid spacing, which limits the spatial dynamic range to 1000 or less on current high-end machines. This makes them useful for simulations of the diffuse intergalactic medium (e.g., \cite{Zhang95}), but is far short of the $10^{4-5}$ dynamic range needed for galaxy large scale structure studies, as discussed above. AMR overcomes this limitation. 

AMR hydrodynamic cosmology codes have been developed by Byran \& Norman \cite{Bryan97, Bryan00}, Kravtsov \cite{Kravtsov99}, and Teyssier 
\cite{Teyssier02}. The Bryan \& Norman code \enzo ~combines a block-SAMR code for ideal gas dynamics using a version of the PPM algorithm adapted to cosmological flows \cite{Bryan95}, with a PM collisionless matter solver as described above (see also paper by O'Shea et al., these proceedings.) The code has been supplemented with the multispecies primordial gas chemistry model of Anninos et al. \cite{Anninos97}, photo-ionization heating by an evolving metagalactic UV background, and a parameterized model for star formation and feedback. The Kravtsov code builds upon the ART N-body code described above, and adds the second-order Godunov solver for ideal gas dynamics described in \cite{Khokhlov98}. The RAMSES code \cite{Teyssier02} developed by Teyssier is similar to the Kravtsov code with minor differences in implementation. 

A problem all three groups have attacked is the formation of an X-ray cluster of galaxies, treating the baryons as non-radiative. This is a good approximation since the $10^8$ K gas characteristic of many X-ray clusters has a cooling time long compared to the Hubble time. A particular set of initial conditions known as the Santa Barbara cluster has served as a community test problem, and is described in Frenk et al. \cite{Frenk99}. The chief difficulty is resolving the X-ray core radius ($\sim$ 100 kpc) in the forming cluster within a simulation volume 64 Mpc on a side. Most of the X-ray luminosity is contained within this region. If one wants to resolve the core radius with 10 cells, say, then a dynamic range of 6,400 is required. Bryan \& Norman \cite{Bryan97} achieved a dynamic range of 8,192 using a base grid of $128^3$ cells and 6 levels of 2x refinement. These results were compiled with those of 11 other codes and presented in \cite{Frenk99}. It was found that to compute the X-ray luminosity to within a factor of 2 accuracy, at least this resolution is required. Kravtsov, Klypin \& Hoffman \cite{Kravtsov02} simulated the Santa Barbara cluster with the ART code with the same resolution as Bryan \& Norman, and found excellent agreement with their results. These simulations, as well as those of Teyssier \cite{Teyssier02}, have shown that the distribution of thermal pressure in the cluster is a much more robust quantity. The significance of this is that Sunyaev-Zeldovich (SZ) effect in clusters of galaxies, which is proportional to the line-of-sight integral of the intracluster gas pressure, is robustly predicted. This makes AMR simulations a powerful tool for guiding upcoming observational surveys of high redshift clusters using the SZ effect (e.g., \cite{Refrigier02}).

A second fruitful application of AMR cosmological hydrodynamics concerns the formation of the first bound objects and stars in the universe. Within the CDM model of structure formation, dark matter begins clustering on small mass scales after the epoch of matter-radiation equality---about 30,000 years after the Big Bang. The characteristic mass scale for DM halos increases with time such that by redshifts of z=20-30, it becomes comparable to the Jeans mass in the expanding, adiabatically cooling, primordial gas. Abel, Bryan \& Norman \cite{ABN00} have used AMR to simulate how baryons collect into the potential well of such a low mass DM halo and the ensuing cooling and contraction of the gas to form a primordial molecular cloud in the halo's center. The simulation used a $64^3$ base grid and 12 levels of 2x refinement for a dynamic range of $2.6 \times 10^5$. In addition to dark matter, gravity and gas dynamics, the calculation solved a 9-species chemical reaction network to model the gas phase reactions which produce molecular hydrogen---the primary coolant in primordial gas. At this resolution, a primordial molecular cloud of size $\sim$ 5 pc was well resolved in a simulation volume 128 kpc (comoving) on a side at z=19. At the end of the calculation, a single, gravitationally unstable cloud core of mass $\sim 100 M_{\odot}$ began collapsing. To follow the evolution of the collapsing core to higher densities, Abel, Bryan \& Norman \cite{ABN02} used an additional 15 levels, for a dynamic range of $10^{10}$. The mesh refinement was driven by the Jeans condition and an analogous condition based on the local cooling time. At this resolution, they were able to refute the prediction by Silk \cite{Silk83} that a chemo-thermal instability would fragment the collapsing core into low mass stars. At the end of the calculation, only a single, collapsing, fully molecular cloud core was found with a size comparable to the Solar System. With mean density of 
$10^{15}$ cm$^{-3}$, this core would trap its cooling radiation and become hydrostatic. Based on the accretion rate at the end of the simulation, it is predicted that the cloud envelope of would accrete in $10^4$ years, forming a Population III star with a mass in the range $30-300 M_{\odot}$. A calculation with 34 levels of refinement (dynamic range $10^{12}$) by Bryan, Abel \& Norman \cite{Bryan01} confirms this result.

\begin{table}
\centering
\begin{tabular}{|l|l|l|}
\hline
\bf{Code}   & \bf{Description} & {\bf URL} \\
\hline
\hline
AMRCLAW &
SAMR infrastructure and hyperbolic solvers &
\cite{AMRCLAW} \\
\hline
AMRCART &
SAMR application: 3D MHD &
\cite{AMRCART} \\
\hline
BEARCLAW &
SAMR infrastructure and PDE solvers &
\cite{BEARCLAW} \\
\hline
CHOMBO &
SAMR infrastructure and PDE solvers &
\cite{CHOMBO} \\
\hline
Enzo &
SAMR application: hydrodynamic cosmology &
\cite{ENZO} \\
\hline
FLASH &
SAMR application: reactive fluid dynamics &
\cite{FLASH} \\
\hline
MLAPM &
AMR application: cosmological N-body &
\cite{MLAPM} \\
\hline
NIRVANA &
SAMR application: 2D and 3D MHD &
\cite{NIRVANA} \\
\hline
PARAMESH &
SAMR infrastructure &
\cite{PARAMESH} \\
\hline
SAMRAI &
SAMR infrastructure &
\cite{SAMRAI} \\
\hline
\hline

\end{tabular}
\newline
\caption{Downloadable AMR Software.}
\label{software}
\end{table}
\medskip

\section{AMR Software}
Here I tabulate some AMR libraries and application codes that are available for download (Table 2). This list is incomplete, because of the lack of centralized information about such tools, as well as the rapid rate of development in the field.

\medskip
\noindent
{\em Acknowledgements:} MLN acknowledges partial support from
the National Computational Science Alliance via NSF Cooperative
Agreement ACI-9619019.

%
%--------------- Bibliography ------------------------------
%
\bibliographystyle{unsrt}
\bibliography{confproceed.bib}
\printindex
\end{document}